# Optical Computing for Deep Neural Network Acceleration: Foundations, Recent Developments, and Emerging Directions

SUDEEP PASRICHA

Department of Electrical and Computer Engineering, Colorado State University, sudeep@colostate.edu

Emerging artificial intelligence applications across the domains of computer vision, natural language processing, graph processing, and sequence prediction increasingly rely on deep neural networks (DNNs). These DNNs require significant compute and memory resources for training and inference. Traditional computing platforms such as CPUs, GPUs, and TPUs are struggling to keep up with the demands of the increasingly complex and diverse DNNs. Optical computing represents an exciting new paradigm for light-speed acceleration of DNN workloads. In this article, we discuss the fundamentals and state-of-the-art developments in optical computing, with an emphasis on DNN acceleration. Various promising approaches are described for engineering optical devices, enhancing optical circuits, and designing architectures that can adapt optical computing to a variety of DNN workloads. Novel techniques for hardware/software co-design that can intelligently tune and map DNN models to improve performance and energy-efficiency on optical computing platforms across high performance and resource constrained embedded, edge, and IoT platforms are also discussed. Lastly, several open problems and future directions for research in this domain are highlighted.

## 1 INTRODUCTION

State-of-the-art artificial intelligence (AI) applications such as ChatGPT and other large language models (LLMs) for natural language processing, object detectors and segmentation models for perception in autonomous vehicles, online search and recommendation engines, protein folding and genomics analysis, and sequence learning for network anomaly prediction rely heavily on a variety of deep neural networks (DNNs) [1]-[6]. These DNN models achieve their high accuracy and performance with the help of massive trainable parameters and large-scale datasets. Training DNN models typically requires heavy computational support in hardware to process problem-specific datasets and update millions of model parameters, for a given learning task. Deploying and using DNN models after training, as part of the inference phase, also requires hardware with the ability to efficiently retrieve and execute the trained DNN models. There are several challenges from a hardware perspective during DNN training and inference. Both training and inference require considerable energy consumption, high memory/storage capacity, and the ability to efficiently support execution of a variety of DNN types, such as Convolutional Neural Networks (CNNs), Recursive Neural Networks (RNNs), Generative Adversarial Networks (GANs), Transformer Neural Networks, and Graph Neural Networks (GNNs).

To accommodate the needs of DNNs, traditional CPU and GPU platforms are rapidly evolving. Intel Xeon Scalable CPUs for instance include the AVX-512 vector processing units for efficient vector-scalar processing required in DNNs [7]. The Tenstorrent Grayskull processor Tenstorrent has a 10x12 array of Tensix cores, each of which is comprised of five RISC cores to provide the thread level parallelism for efficient vector processing [8]. Nvidia GPUs such as the Volta V100 and subsequent generations of architectures such as Ampere and Lovelace include Tensor cores designed specifically to improve operations on Tensor (high-dimensional array) data in DNNs [9].

Beyond changes in these traditional general-purpose processing platforms, hardware architects have been finding creative ways to meet the growing computational demands of DNNs, with innovations in specialized circuits and chip design to accelerate often-used operational kernels, methods, and functions in DNNs. Many such DNN accelerators have emerged over the past decade, including various digital ASIC designs (Google's TPUs ver. 1-4, Intel Movidius, Mythic Intelligent Processing Unit) [10]-[12], analog and mixed-signal neuromorphic processors (IBM TrueNorth, Intel's Loihi) [13], [14], and FPGA-based offerings (Intel Arria and Microsoft Brainwave) [15], [16]. Today, many System-on-Chip (SoC) platforms used in embedded and IoT applications integrate custom processing cores for DNN acceleration, e.g., neural engines within Apple's A17 Pro chip found in the iPhone 15 Pro and Pro Max [17]. The capabilities of these DNN accelerators vary considerably, with different accelerators balancing performance, energy efficiency, size/weight, and functional flexibility in unique ways. Often, accelerators target either DNN training or inference [18]. Training accelerators have larger power budgets of 100W or higher and support for greater numerical precision, such as 16-bit floating point numbers (fp16 and bfloat16). Inference accelerators have smaller power budgets of less than 100W and support for smaller numerical precision, such as at most 8-bit integer (int8) operands. These inference accelerators can be further categorized into ultra low power accelerators for applications such as speech processing and lightweight sensing with a less than 1W power budget, embedded accelerators for applications such as wireless cameras, small UAVs, and robots with a less than 10W budget, and autonomous accelerators for driver assistance services, autonomous driving, and autonomous robots with a less than 100W budget.

Increasingly, hardware architects are realizing that it is becoming harder to achieve scalable performance with CMOS-based electronic general-purpose and domain-specific DNN accelerators, due to the ending of Moore's law and related trends, such as Denard's scaling (power density) and Koomey's law (instructions per Joule) [19]. Addressing this scalability problem is a critical issue to avoid a second winter in the AI industry in the near future. Fortunately, integrated optical computing is emerging as a promising solution to address this problem. Optical computation involves communicating and processing information with light signals and with limited reliance on slower electronic signals [20]. The optical computing paradigm offers many compelling advantages, such as throughput rates of hundreds of terabits per second, enhanced power efficiency due to minimal heat generation from optical signals, long-distance transmission with minimal energy loss, and the inherent parallelism of light, which can concurrently traverse multiple paths to support multiple operations in parallel at low power requirements, on the order of a few watts per teraflop [21]. This article discusses the evolution, state-of-the-art, challenges, and future directions with optical computing for DNN acceleration.

The rest of this article is organized as follows. Section 2 discusses the history of computing in the optical domain and the emergence of integrated optical computing. Section 3 presents the fundamentals and building blocks of integrated optical computing. Section 4 discusses examples of cross-layer design of optical computing platforms for DNN acceleration. Section 5 discusses examples of hardware/software codesign of optical computing platforms for DNN acceleration on resource limited edge, IoT, and embedded applications. Section 6 discusses open problems and future directions with optical computing for DNN acceleration. Lastly, Section 7 concludes this article.

## 2 HISTORY OF OPTICAL COMPUTING

The origins of modern optical computing date back to 1953 when one of the earliest works proposed the idea of exploiting the speed and parallelism of light signals to process information at high data rates for real-time pattern recognition [22]. The architecture of this early tabletop optical processor consisted of three parts: the input, the processing plane, and the output plane. The input consisted of the data to be processed, which initially was a fixed image slide, but was later replaced with a spatial light modulator (SLM) for electrical to optical signal conversion. The processing plane consisted of lenses and utilized their ability to perform Fourier transforms on the amplitude and phase of an input signal. Later, nonlinear components and holograms were also explored as part of this plane. The output plane was composed of photodetector array or a camera to detect the results of the processing. This simple architecture was extended between the 1950s and 1970s for real-time spatial filtering, character recognition, and correlation analysis [23]-[25]. As an example, a joint transform correlator (JTC) was proposed in [25]. Two images (a reference and a scene) were placed adjacent to each other in the input plane and a Fourier transform was performed on both of them by a first lens. The intensity of this joint spectrum was detected by a CCD camera and then a Fourier transform was performed optically using a second lens, which generated the correlation between the input scene and the reference in the output plane. Figure 1(a) shows the setup from [25] which generated the crosscorrelations between the scene and the reference. Figure 1(b) depicts the output plane of the JTC for the case when the reference and the scene are identical, indicated by the two similar crosscorrelation peaks. Such a system was used for real-time target tracking. Note that this system is a type of coherent optical processor where information is carried by complex light wave amplitudes [26]. In contrast, noncoherent (aka incoherent) optical processors utilized the intensities of light waves as information carriers. Such noncoherent optical processors had the desirable properties of not being sensitive to phase variations in the input plane and exhibiting no coherent noise, and were used in various signal processing applications [27]. Both coherent and noncoherent systems excelled at linear optical processing of space invariant operations such as correlation and convolution, or space-variant operations such as coordinates transforms and Hough transforms [28]. It is also possible to optically implement nonlinear processing such as logarithm transformation, thresholding, or analog to digital conversion [29].

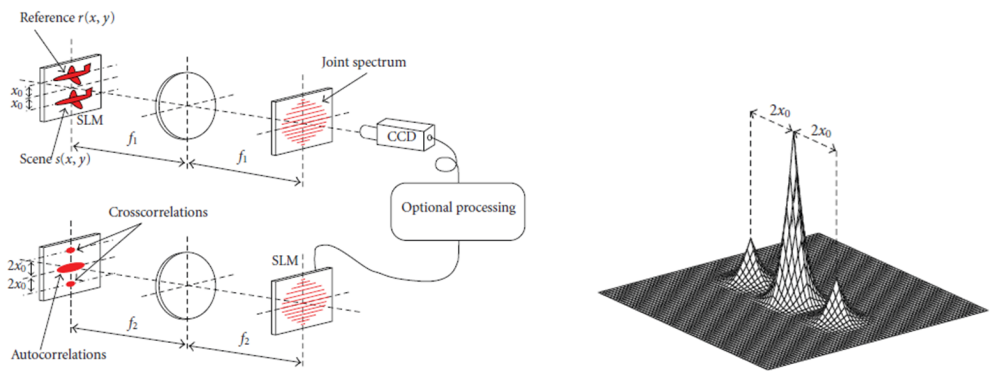

Figure 1. Joint transform correlator (JTC): (a) Optical setup. (b) Output plane of the JTC. From [30]

Between the 1970s and 2000s, there was significant and growing interest in optical computing. Several efforts improved the design of the SLM, which was a key component in optical processing to control the amplitude and the phase modulation of the input or the filter plane [31]. A variety of SLM designs were proposed and prototyped during this period, including liquid crystal SLMs, multiple quantum wells devices (MQW), magneto-optic SLMs, and Deformable Mirror Devices (DMDs) [32]. These newer SLMs were used to design many optical processing systems in laboratories, including for detecting tracks in a high-energy physics experiment [33], performing Hough transforms [34], real-time road sign recognition [35], and automatic target recognition [36]. While several of these optical processors were made available commercially, they were not very successful due to high costs and niche application focus. However, today, SLMs based on liquid crystals and DMDs (using MEMs based implementations) have found commercial success in optical communication systems for signal modulation, in display devices such as liquid crystal microdisplays and holographic displays, as well as to a limited extent across applications such as pulse shaping, quantum key distribution, and optical metrology [37]-[39].

Around the 1990s and early 2000s, interest in optical computing started to decline. The main reason was the proliferation of digital computers which were more flexible and general-purpose than optical processing systems that were often designed to accomplish a very specific processing task. Not only were digital computers improving in terms of performance and energy efficiency, but they were also cheaper, smaller, and more easily accessible to non-experts than bulkier, expensive, and niche optical processing solutions. The optical processing systems discussed so far were primarily analog, but some efforts also emerged to realize digital optical computers that could compete more efficiently with digital electronic computers. For instance, a 32-bit, fully programmable digital optical computer (DOC II) designed to operate in a UNIX environment running RISC microcode was proposed in [40]. However, such systems could not compete with digital electronic computers that were rapidly being miniaturized, made more cheaply, and achieved far greater performance due to advances in semiconductor technology scaling than was possible with digital optical computing.

While optical computing was in decline, an interesting development was occurring in the 2000s that would have significant implications. There emerged growing interest in realizing photonic integrated circuits on silicon chips, by relying on the developments in integrated optics that had led to the proliferation of large-scale optical communication networks in the telecommunication domain since the 1980s. The term integrated photonics refers to the fabrication and integration of multiple photonic components such as couplers, beam splitters, gratings, polarizers, and interferometers on a common planar substrate and interconnecting them via optical transmission lines called waveguides [41]. The integrated optics solutions that emerged starting around the 1980s often included elements for the generation, focusing, splitting, junction, coupling, isolation, polarization control, switching, modulation, filtering and light detection on the same chip. Such integration of multiple optical functions on a single photonic device was a key step towards miniaturizing optical systems and lowering the costs of advanced optical systems. The development and integration of novel optical devices enabled the integration of sophisticated functions such as Wavelength Division Multiplexing (WDM; the transmission of several light signals through a single optical fiber using several wavelengths) which was crucial to meet the growing bandwidth requirements of telecommunication domain optical communication networks that needed to support growing data traffic from the Internet.

Initial interest in realizing photonic integrated circuits on silicon chips in the 2000s was limited to overcoming data movement challenges that plagued electronic chips [42]. The idea was to use photons rather than electrons to transfer data and signals at the chip-scale, due to the high frequency of light (~200 THz), which enables a very large bandwidth for transporting a large amount of information. Such optical communication was seen as a means to overcome the physical limitations of electronic conductors that were beginning to suffer from longer signal propagation delays and increasing interference with shrinking fabrication technology nodes. Although lithographic fabrication of optical devices requires some materials that are different than those used in traditional electronics, the processes are essentially the same, and the techniques that have been well established from decades of semiconductor production can be used. A CMOS-compatible lithographic system for fabricating optical components can use more or less the same set of tools as in electronics, including photoresists, exposure tools, masks, and all the pattern transfer process from mask to resist and then to device [41]. The standardization of silicon photonic integrated circuits over the past few decades has led to the adoption of photonic links at smaller and smaller scales. For instance, silicon photonic transceivers are now a pervasive component in datacenters to enable optical connections between servers, storage nodes, and network switches [43]. Companies such as Intel started exploring the possibilities of replacing electrical interconnects by optical interconnects at the even smaller chip scales starting in the mid 2000s, due to the terahertz bandwidth, low loss, and low crosstalk with photonics [44], [45]. Several interesting optical bus-based and network-on-chip (NoC) interconnect architectures were proposed at the chip-scale to overcome data movement bottlenecks in multicore computing systems [46]-[49]. These developments extended into the 2010s, with several innovative optical communication architectures being proposed at the chip-scale that addressed bandwidth, quality-of-service, reliability, and security objectives for various embedded and high performance applications [50]-[63].

The resurgence of AI in the form of deep neural networks (DNNs) in the 2010s proved to be a fortuitous development that spurred interest in optical computing. Due to the growing computational complexity of deep neural networks (DNNs) that required millions (sometimes billions) of matrix-vector multiplications and additions during both training and inference, domain specific hardware to accelerate AI applications began to be investigated. But as the pace of growth with DNNs (approximately doubling in size every 3.5 months [64]) is far outpacing Moore's law, traditional electronic DNN accelerators are expected to lag behind the demands of such models in the near future. One of the promising directions that is now possible to explore due to the maturation of photonic integrated circuits on silicon is the ability to perform optical computing on silicon to facilitate DNN computations, with higher compute density, bandwidth, cost efficiency, and energy efficiency than was ever possible with non-integrated optical computing platforms in the past [20].

Current DNN models require massive numbers of dense, low-precision matrix computations such as multiplications and additions. Electronically performing these operations suffers from high communication overheads and high latency digital operations. In contrast, photonic matrix operations can be performed passively, with much lower energy scaling costs of $O(N)$ for $O(N^2)$ fixed point operations [65]. Operations such as photonic matrix multiplications can occur in a single step, limited only by the modulation and detection steps at the periphery. Moreover, despite the larger sizes of photonic devices compared to electronic transistors and logic gates, photonic systems can deliver more operations per second in a given area than digital electronic systems [66]. Lastly, DNNs require relatively long-range connections between neurons and neuron layers, to perform distributed information processing. Compared to metal wires, optical signals generate less heat, experience lower attenuation, and utilize waveguides that have no inductance or skin effect, which means that frequency-dependent signal distortions are minimal for the long-range connections present in DNNs [67]. In the next section, we discuss the fundamentals of integrated optical computing for DNN acceleration.

## 3 FUNDAMENTALS OF INTEGRATED OPTICAL COMPUTING FOR DNN ACCELERATION

The idea of accelerating DNN applications with optical computing is not new. In the 1980s, optical processors were designed for matrix operations [68], systolic array processing [69] and neural network processing [70]. However, these processors were bulky, inflexible, and not integrated into the silicon ecosystem that is the foundation of today's compact processing platforms. State-of-the-art deep learning models such as DNNs can learn complex nonlinear functions by cascading layers of linear matrix operations with non-linear activation functions. Matrix operations, particularly matrix multiplications, tend to dominate the computation time when executing DNNs, e.g., taking more than 90% of the time [71]. Integrated silicon photonic neural network implementations of matrix operations can significantly accelerate DNN model execution. In the following subsections, we discuss the differences between analog and digital computing, building blocks of optical computing systems, and types of integrated optical neural network implementations based on light coherence principles.

### 3.1 Digital vs. Analog Processing

Traditional digital electronics based on the von Neumann architecture has been dominant for the past 60 years, driven by the rapid growth dictated by Moore's law (the number of transistors that can be put on a chip doubles every 18 to 24 months) and Koomey's law (the number of computations per joule of energy dissipated doubles approximately every 1.57 years) [72]. But these architectures also suffer from the well-known von Neumann bottleneck [73] which fundamentally limits the performance of these systems especially when executing memory-intensive workloads such as DNNs. To overcome these limitations, architects have favored massive parallelization and specialization of processor architectures to improve DNN performance. However, several barriers to scale processing density and energy density remain: 1) With clock rates saturating to around 4 GHz, architects have increasingly focused on performance gains from parallelization, which tend to lead to diminishing returns as predicted by Amdahl's law [74], 2) The increasing dynamic and leakage power overheads of parallel architectures in deep nanometer semiconductor technology regimes leads to greater heat production and thermal effects that in turn create stringent limits on chip scalability [75], 3) Signal bandwidth on metallic interconnects is limited to a few gigahertz in frequency due to harsh trade-offs between bandwidth and interconnectivity [76], 4) The energy for data movement over metal wires can also exceed half the total energy in specialized processors [19], and 5) The overall energy efficiency of around 1 pJ per MAC (multiply and accumulate computation) is six orders of magnitude higher than for biological systems that are estimated to consume 1 aJ per MAC [77].

Analog processing in the photonic domain can achieve much higher bandwidth densities and also consume less energy for longer distances than electrical counterparts [78]. The advantages of photonics are particularly potent for systems with high communication or bandwidth requirements [66]. Optical signals (carrier waves) possess different orthogonal features, such as wavelength, spatial mode, and polarization, which do not interact with each other in passive devices and create a complex electric field that is a sum over each feature. The optical communication band has an approximately 5 THz spectral bandwidth (for wavelength-division-multiplexed systems in the 1300 nm or 1550 nm wavelength bands), which provides around 5 Tb/s of information capacity across each mode and polarization. Unlike electronic links, this bandwidth is an intrinsic property of the electromagnetic wave, is far superior to what is possible in electronic links, and is independent of physical design constraints such as waveguide length or proximity. Many of the design constraints that limit microwave electronic circuits are much less of an issue in optical systems, where it is sufficient to match the refractive index to prevent reflections. As electro/optical (E/O) and opto/electrical (O/E) conversion is an inherently quantum process, electrical nodes that communicate using photonics need not be electrically impedance matched with one another [79]. Further, as photonic signals are not subject to Joule heating, it is possible to achieve very low signal attenuation with waveguides and fibers, which results in communication costs that scale almost independently of distance. Not only does this allow propagation of higher power signals without any associated thermal runaway concerns, but communication or computations in the optical domain can be performed with (theoretically) near zero energy consumption, especially for linear or unitary operations.

It should be noted that analog operations are far more resolution limited than standard floating-point operations in digital electronic systems. As analog systems use physical representations of real numbers, they lack the dynamic range to represent different exponents [66]. Therefore, analog operations are restricted to fixed point operations, in which the exponent is fixed during computation. Fortunately, extensive research has shown that DNNs can operate effectively with both low precision and fixed-point operations. DNN inference accuracy has been shown to negligibly impacted with 4-8 bits of precision in both activations and weights (and sometimes even lower than that [80]), while 8–16 bits of precision per computation can suffice for model training [81]. Thus, analog photonic processing is a good fit for DNN processing as it has been shown to exhibit a tuning accuracy of more than 4 bits [65], [82].

### 3.2 Building Block Components

There are several fundamental photonic and optoelectronic devices and circuits used for optical computing.

Lasers (either off-chip or on-chip) are used to generate optical signals that are required to perform computation and communication in optical circuits. Off-chip lasers have high light-emitting efficiency and good temperature stability at a cost of large coupling losses (which leads to larger energy consumption) between the off-chip light source and the silicon chip, which is mostly due to the grating coupler loss, as well as higher packaging costs. In contrast, on-chip lasers can potentially achieve a higher integration density and better performance in terms of energy efficiency, but the development of on-chip lasers on silicon is extremely challenging because of the low emission efficiency of silicon, and its sensitivity to chip-wide thermal variations [83]. As photonic processing systems do not typically need to send optical signals off-chip, ideally, all optical signals would be confined within an integrated circuit package. Thus, co-packaged laser sources are critical for the efficiency, stability, and scalability of photonic processing.

Waveguides are the electrical wire counterparts in integrated silicon photonic circuits and carry the optical signal(s) generated by the laser source. They are composed of two materials, resulting in a high-refractive-index contrast. Typically, in silicon photonics, these materials are Silicon (for the waveguide core) and Silicon Dioxide (for the waveguide cladding). This combination of core/cladding materials enables total internal reflection, allowing optical signals to stay confined within the waveguide. The waveguides can either have a ridge or strip shape [84]. Wavelength Division Multiplexing (WDM) allows a single waveguide to support multiple signal carrier wavelengths simultaneously without any interference. This allows for the transmission of ultra-high bandwidth signals and is useful for performing MAC operations with high throughput. One challenge with optical signal propagation in waveguides is that these signals experience optical loss (i.e., the propagation loss, typically characterized in dB/cm) due to various factors, such as some imperfections in the waveguide structure (e.g., waveguide sidewall roughness). In addition to propagation loss, waveguide bends lead to optical bending loss where an optical signal will be attenuated due to the radiation loss and mode-mismatch in waveguide bends. This bending loss is proportional to the radius of the waveguide bend. Limiting optical loss in silicon photonic waveguides is essential as it limits the scalability of photonic processing and substantially degrades the energy efficiency in such systems.

Modulator devices placed along waveguides are used to couple (inject), remove (filter out), and manipulate (alter) optical signals in the waveguide. Mach-Zehnder Interferometers (MZIs) [85] and microring resonators (MRR) [86] are the two most common silicon photonic devices used for modulation. MZIs consist of two waveguide arms with directional couplers and phase shifters. The phase shifters use tuning circuits to change the optical phase in one or both arms of the MZI, introducing constructive or destructive interferences at the output, which allows switching an optical signal between the output ports. MRRs consist of a ring-shaped waveguide and are specifically designed and adjusted with the help of tuning circuits to couple, remove, or manipulate a particular wavelength, known as the MRR resonant wavelength. These modulators are the key components that directly perform photonic processing and facilitate photonic communication.

Tuning circuits are integrated with modulator devices and are designed to control their behavior. These circuits allow precisely modifying characteristics (e.g., phase, amplitude) of an optical signal passing in the vicinity of these devices. Typically, the tuning circuit can be one of two types: carrier injection Electro-Optic (EO) tuning or Thermo-Optic (TO) tuning. EO tuning can provide a tuning range of only 1.5 nm at most, but it incurs relatively low latency and power overheads [87]. In contrast, thermal tuning incurs high latency and power overheads, but it can provide a larger tuning range of about 6.6 nm and induces lower power loss than current injection tuning [56]. It is possible to only rely on one of these tuning methods in a modulator design, but intelligently utilizing both can enable better energy efficiency.

Photodetectors (PDs) are needed to detect processed optical signals and convert them into electrical signals. Typically, a modulator filter device (such as an MRR) couples light out of a waveguide and drops it on a PD to generate a proportional electrical current. To be effective, a PD should be able to generate an electrical output even for a small input optical signal. The input signal power from the laser source must be greater than the responsivity of the PD. Calculating the required input signal power for correct operation requires taking into consideration the different types of losses that may occur along the optical link, and ensuring that the power exceeds the responsivity constraint of the PD at the destination.

Digital-to-Analog Converters (DACs) and Analog-to-Digital Converters (ADCs) are used to convert digital information (0s and 1s) into a proportional electrical current, and to convert the electrical current generated by a PD into digital information. These devices represent one of the main performance bottlenecks in silicon-photonic-based processing systems due to their high latency and power costs.

### 3.3 Noncoherent vs. Coherent Neuron Implementations

The building blocks discussed in the previous subsection are combined to implement the neurons within DNNs in the photonic hardware. A photonically implemented neuron is analogous to an artificial neuron and consists of three components: a weighting, a summation, and a nonlinear unit. The neuron performs a multiplication between the incoming activations with scalar weights, aggregates (sums) the resulting products, and applies a nonlinear transformation to generate an output. In a DNN, the output of each neuron in a layer is typically connected to downstream neurons in the next layer. There are two popular approaches that have been adopted for implementing photonic neurons: noncoherent and coherent.

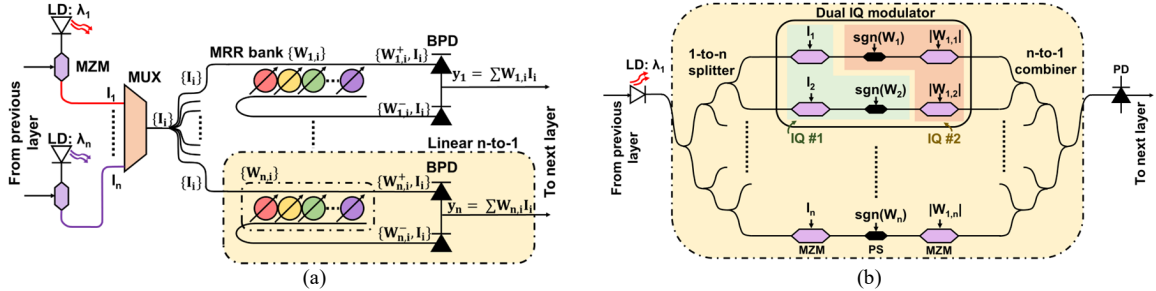

Figure 2. Two main types of photonic neurons: (a) noncoherent Broadcast-and-weight optical neurons [88], (b) coherent optical linear neuron [89]. (MRR: micro-ring resonator; LD: laser diode; MZM: Mach–Zehnder modulator; BPD: balanced photodetector; PS: phase shifter; PD: photodiode; IQ: in-phase-and-quadrature; OLAU: optical linear algebra unit).

Noncoherent neuron implementations rely on manipulating optical signal power (intensity) across multiple wavelengths. Figure 2(a) shows n neurons in a layer, with the dotted box representing a single noncoherent neuron [88]. Each input (activation) to a neuron is imprinted using a Mach–Zehnder modulator (MZM) onto a unique wavelength ($\lambda_i$) emitted by a laser diode (LD). The wavelengths are multiplexed (MUXed) into a single waveguide (i.e., WDM) and split into n branches that are each weighed with a micro-ring resonator (MRR) bank that alters optical signal power proportional to weight values. A balanced photodetector (BPD) performs summation across positive and negative weight arms at each branch. One of the earliest noncoherent optical implementations for neural network acceleration was discussed in a 2014 paper [90] and demonstrated in 2017 [88]. Since then, several other noncoherent architectures have been proposed that employ WDM optical signal fan-in for weighted addition [91]-[93], with differences between them in how channels are weighted and the type of devices used for weighting, e.g., using tunable attenuators designed with semiconductor optical amplifiers (SOAs) or phase change materials (PCMs). Another approach to optical domain weighing involves modulating the effective refractive index of a waveguide carrying optical signals, e.g., using waveguide-embedded heaters [94]. The integration of PCMs within tuning methods can allow weights to retain their values without further power dissipation after being set [95].

Coherent neuron implementations manipulate the electrical field amplitude and phase with a single wavelength. They rely on destructive or constructive interference effects in MZM devices (which are based on a Mach-Zehnder interferometer (MZI) devices that split light into two branches and then recombine them by interference). Figure 2(b) shows an example of a single coherent neuron [89]. A modulator within it works on two inputs at a time, imprinting input activations with MZMs on the two branches. Weighing occurs with electrical field amplitude attenuation proportional to the weight magnitude, and phase modulation that is proportional to the sign of the weight. The weighted signals are accumulated with cascaded Y-junction combiners and sent to a photodiode (PD) to obtain an electrical output. An architecture with arrays of beam splitters and phase shifters was described in an early work from 1994 [96], which was able to implement unitary matrix transforms using interference between different paths of coherent input light (with inputs assigned to different waveguides and power modulated). An implementation of this unitary transformation architecture was shown in 2015 [97], with an integrated platform that included thermally tuned silicon waveguides and directional couplers arranged in a mesh of MZIs. As neural networks require computations with matrices that may not necessarily be unitary, this work was extended in 2017 [65] in a prototype that factored a weight matrix into one unitary MZI-mesh, one array of tunable attenuators, and a second unitary MZI-mesh.

Multiply and accumulate (MAC) operations in the photonic domain can be performed at very high speeds with the implementation discussed above, limited only by the optoelectronic devices that encode and decode the signals at the input and output [66]. Considering typical bandwidths of greater than 20 GHz per photonic device, the signal bandwidth of each input can readily exceed 10 GS/s for a given NxN matrix operation. Further, considering a less than 50 ps delay for most photonic components results in a delay that is approximately 100 ps for the same matrix operation. Thus, the entire matrix can be computed in less than a single digital electronic clock cycle. These numbers are orders of magnitude better than the~µs latencies and greater than 1 ns speeds observed with electronic approaches [98].

Beyond multiplication and summation, some form of nonlinearity is required in the primary signal pathway to implement the thresholding effect of a neuron. Photonic neurons must be capable of reacting to multiple optical inputs (fan-in), apply a nonlinearity, and producing an optical output suitable to drive other photonic neurons. For both noncoherent and coherent photonic neuron implementations, non-linearity can be induced with two main categories of devices: optical-electrical-optical (O/E/O) and all-optical. In O/E/O device based neurons, a transduction of optical power into electrical current and back is realized within the primary signal pathway. Here nonlinearities can occur in the electronic domain or in the E/O conversion stage using lasers or saturated modulators [99]. In contrast, all-optical device based neurons do not ever represent the neuron signal as an electrical current but rather as changes in material properties, e.g., semiconductor carriers or optical susceptibility. But optical nonlinear susceptibilities are generally power inefficient which means that the output is significantly weaker than its input. This can hurt cascadability of photonic neurons, as the weak output may be incapable of driving another neuron. To overcome this challenge, nonlinear optical devices can be combined with optical carrier regeneration approaches. Here regeneration involves generating a fresh carrier wave, which is power modulated by a neuron's output signal to overcome the cascadability limits of all-optical nonlinear devices. Various approaches for such regeneration are possible, for instance using cross-gain modulation or cross-phase modulation in an interferometer [100] and optical amplifiers to boost the output such that it can drive downstream neurons [101].

There are some significant differences between coherent and noncoherent optical computing discussed here. The footprint of an MZI in coherent implementations can be over 10,000μm$^2$, which results in a very low computing density in coherent MZI based implementations of neuron layers. MRR based noncoherent implementations are much more compact, as MRR devices possess a diameter of only a few microns. Coherent implementations must also grapple with both amplitude and phase manipulation, and it is very difficult to prevent phase noise accumulation from one nonlinear stage to another. Further, coherent implementations are limited to only one wavelength (otherwise constructive and destructive interference will not occur between interacting light waves) which limits the computational throughput, unlike noncoherent implementations that can have more inherent parallelism and throughput as they operate on multiple wavelengths simultaneously. For these reasons, noncoherent implementations are more promising.

## 4 CROSS LAYER DESIGN OF OPTICAL COMPUTING FOR DNN ACCELERATION

In this section, we discuss the various challenges with photonic acceleration of AI applications (specifically DNNs) and describe various solutions to overcome these challenges. The focus is on noncoherent optical computing, due its advantages over coherent optical computing, as discussed in the previous section.

### 4.1 Implementation Challenges

There are several challenges with designing a noncoherent photonic accelerator for DNNs:

- **Variations:** The reliability of optical computing can be adversely impacted by semiconductor fabrication process variations and thermal variations. Process variations are the undesirable but uncontrollable variations during device manufacture, which can cause the devices to behave sub-optimally. For example, experimental studies have shown that MRR devices (discussed in Section 3.2) used in optical computing implementations can experience significant resonant drifts (~9 nm reported in [102]) within a wafer due to process variations. This matters because even a small 0.25 nm drift can cause the bit-error-rate (BER) of photonic data traversal to degrade from 10-12 to 10-6. Such errors lead to erroneous data at the receiver of a photonic link, or incorrect outputs with optical computation that may not be correctable with available error correction techniques. Variations in temperature during DNN execution can also cause resonant drifts and create thermal crosstalk in devices such as MRRs, which can reduce the achievable precision (i.e., resolution) of weight and bias parameters. This in turn can reduce DNN model accuracy when these models are executed on the variation-susceptible optical computing platforms.
- **Tuning:** As discussed in Section 3.2, tuning circuits are essential in active photonic devices such as MRRs to enable error-free computational operations and modulation/filtering. Typically, these circuits can be based on either thermo-optic (TO) or electro-optic (EO) principles. While EO tuning is faster (~ns range) and consumes low power (~few μW), it suffers from a smaller tuning range. The smaller tuning range implies that in certain scenarios, EO tuning may not be able to correctly tune a device. In contrast, TO tuning has a larger tunability range that can ensure correct tuning of devices in all practical scenarios but consumes higher power (~few mW) and has a much higher (~μs range) latency. Thus, the overheads and constraints of tuning mechanisms need to be carefully considered in the design of optical computing platforms.
- **Losses:** Optical signals traversing waveguides encounter losses in signal intensity, such as propagation loss (often characterized in dB/cm) due to some imperfections in the waveguide structure, e.g., waveguide sidewall roughness. In addition to propagation loss, many other types of losses also exist. Waveguide bends create optical bending losses (proportional to the radius of the waveguide bend) where an optical signal is attenuated due to the mode-mismatch and radiation loss in waveguide bends. Coupling laser signals from optical fibers into on-chip waveguides can create coupling losses due to mismatches in waveguide dimensions. The imperfect nature of modulation with devices such as MRRs can lead to modulation loss as well as through loss, where the latter occurs due to the undesirable coupling of optical signals with off-resonance MRR devices in the vicinity of the waveguide that the signals are traversing. Losses also occur when splitter/combiner devices are used to split or merge signals across multiple waveguides, and during EO/TO tuning. Minimizing such optical losses is essential as it limits the scalability of optically implemented DNNs and substantially degrades the power and energy efficiency in such platforms (as losses must be compensated by increasing laser power [103]).

- **Optical-Electrical Conversions:** Conversions from the electrical to the optical domain and back are inevitable, given that DNNs require access to activations and parameters that can be most efficiently stored in digital memories. Moreover, certain complex operations may not be possible in the optical domain and may require digital circuits. For the conversions, we need to make use of analog-to-digital converters (ADCs) and digital-to-analog converters (DACs). These DAC/ADC devices have relatively high energy consumption overheads [20]. Thus, there is a need to efficiently design optical computing architectures that minimize the need for a large number of these conversion devices, to improve the overall energy efficiency of DNN execution.

- **Mapping:** The way in which DNN computations are decomposed, orchestrated, and ultimately mapped to the various optical devices is crucial to achieving high performance and energy efficiency in optical computing. In DNN models, the matrix operations can be decomposed into vector-dot-product operations. It is important to efficiently perform these decomposition and mapping steps, so that vector dot products can occur without bottlenecks. This requires careful co-design of the decomposing/mapping strategies in software and the optical computing architecture implementation in hardware.

## 4.2 Accelerating CNNs

Convolutional neural networks (CNNs) are examples of DNN algorithms that have been successfully used for a variety of computer vision problems, such as image classification, semantic segmentation, object detection, and pose estimation. In [104], a novel cross-layer optimized noncoherent optical computing accelerator targeting CNNs was presented. The proposed CrossLight accelerator featured multiple innovations across design layers to overcome many of the challenges highlighted in Section 4.1. At the device level, a comprehensive design-space exploration of MRRs was performed to compensate for fabrication process variations. This exploration led to the insight that that in an MRR design of any radii and gap, when the input waveguide is 400 nm wide and the ring waveguide is 800 nm wide at room temperature (300 K), the undesired resonant wavelength drift due to fabrication process variations can be reduced from 7.1 to 2.1 nm (70% reduction). This was a significant result, as these engineered MRRs require less compensation for variation-induced resonant wavelength drifts, which can reduce the power consumption of architectures using such MRRs. Additionally, to overcome thermal crosstalk, optimized placement of devices was proposed. At the circuit level, a hybrid tuning approach was proposed to reduce reliance on the power hungry and slow TO tuning. The hybrid approach opportunistically used faster and more power efficient EO tuning, and only resorted to using TO tuning infrequently, when larger resonant drifts needed to be compensated. The approach also reduced thermal crosstalk from TO tuners, which led to more compact layouts. At the architecture level, vector dot product (VDP) units were designed, consisting of banks (groups) of MRRs to imprint both activations and weights onto the optical signal. Two different types of VDPs were designed: one to support convolution (CONV) layer acceleration and the other to support fully connected (FC) layer acceleration. Both of these layers are extensively used in CNNs. For CONV layer acceleration, n VDP units were designed, with each unit supporting an N×N dot product. For FC layer acceleration, m units were designed, with each unit supporting a K×K dot product. Here n>m and K>N, as per the unique requirements of each of the distinct layers. In each of the VDP units, the original vector dimensions were decomposed into N or K dimensional vectors. Lastly, the architecture was designed to promote significant reuse of wavelengths in each VDP, to minimize laser power.

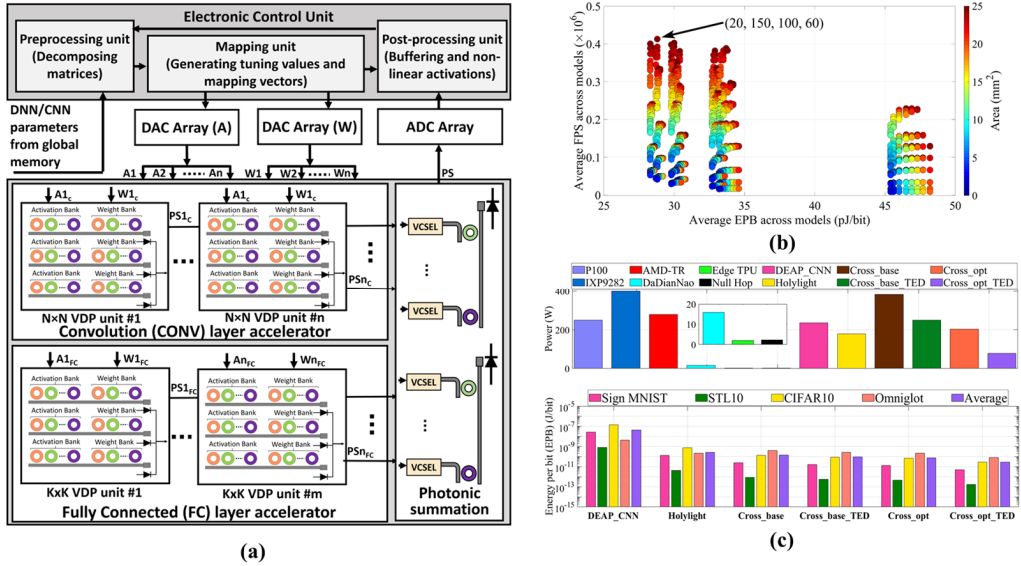

Figure 3. (a) An overview of CrossLight optical CNN accelerator, showing dedicated vector dot product (VDP) units for CONV and FC layer acceleration, (b) Scatterplot of avg. FPS vs. avg. EPB vs. area of various CrossLight configurations; the configuration with highest FPS/EPB and FPS is highlighted, (c) Power consumption comparison among variants of CrossLight vs. optical accelerators (DEAP-CNN, Holylight), and electronic accelerators (P100, Xeon 9282, Threadripper 3970x, DaDianNao, EdgeTPU, Null Hop), as well as energy comparison of EPB values of the optical CNN accelerators. From [104].

Figure 3(a) shows a high-level overview of the CrossLight optical accelerator substrate that performs vector dot product operations using silicon photonic MRR devices, and summation using balanced photodetector (PD) devices over multiple wavelengths. An electronic control unit controls the photonic devices, and orchestrates communication with a global memory to obtain the parameter values, perform mapping of the vectors, and buffer partial sums. DAC arrays are used to convert buffered signals into analog tuning signals for MRRs. ADC arrays are used to map the output analog signals from PDs to digital values that are sent back for post-processing and buffering. Figure 3(b) shows the results of an exploratory analysis performed by varying the number of VDP units for the CONV layers (*n*) and FC layers (*m*), along with the complexity of the VDP units (*N* and *K*, respectively). The results shown indicate the frames per second (FPS; a measure of DNN inference performance) vs. energy per bit (EPB) vs. area of various configurations of *CrossLight*. The best configuration that was selected had the highest value of FPS/EPB, and in terms of [*N, K, n, m*], the values of the four parameters for this configuration are [20, 150, 100, 60].

Figure 3(c) shows the power and EPB comparison of CrossLight against two well-known optical CNN accelerators (DEAP-CNN and Holylight; with a similar area constraint for all accelerators of ~16-25 mm$^2$) as well as electronic platforms: three deep learning accelerators (DaDianNao, Null Hop, and EdgeTPU), a GPU (Nvidia Tesla P100), and CPUs (Intel Xeon Platinum 9282 denoted as IXP9282, and AMD Threadripper 3970x denoted as AMD-TR). Results for four variants of the CrossLight architecture are shown: 1) *Cross_base* utilizes conventional MRR designs (without fabrication process variation resilience) and traditional TO tuning; 2) *Cross_opt* utilizes the optimized MRR designs, and traditional TO tuning; 3) *Cross_base_TED* utilizes the conventional MRR designs with the hybrid tuning approach discussed above; and 4) *Cross_opt_TED* utilizes the optimized MRR designs and the hybrid tuning approach. It can be observed that on average, the best CrossLight configuration (*Cross_opt_TED*) has lower power consumption than both optical accelerators, as well as the CPU and GPU platforms, although this power is higher than that of the edge/mobile electronic accelerators. Further, this configuration has 1544× and 9.5× lower EPB compared to DEAP-CNN and Holylight, respectively. The reason for the significantly lower EPB with CrossLight is because it considers and optimizes various losses and crosstalk that an optical DNN accelerator would experience, and puts in place novel cross-layer approaches at the device, circuit, and architecture layers to counteract their impact.

### 4.3 Accelerating RNNs

Recurrent Neural Networks (RNNs) are a class of DNNs that possess internal memory and feedback connections that make them well suited for learning trends and patterns found in sequences (e.g., time series data) where the data elements are correlated. As a result, RNNs have been found to perform well for sequence learning tasks, such as continuous human activity recognition, temperature prediction based on time series sensor data, etc. But simple RNNs are ineffective when forced to learn large sequences of data, due to the problem of vanishing gradients. To overcome this problem, more advanced RNN models have been proposed, which include more sophisticated neuron implementations: Gated Recurrent Units (GRUs) and Long Short-Term Memory (LSTM). Compared to simple RNNs, the gates and states used in GRUs and LSTMs make them much more effective for learning long-term dependencies.

Figure 4(a) shows a simple RNN cell at a timestep *t* consisting of an input ($x_t$), an output ($o_t$), and a hidden state ($h_t$) that allows it to remember information learned from a sequence. The simple RNN cell uses a *tanh* activation function to compute the hidden state. This hidden state holds information about the inputs from the previous timesteps. The current stage output of a simple RNN is computed by taking the previous hidden-state information ($h_{t-1}$) along with the input. Compared to the simple RNN cell that can only learn short term trends in sequences, GRU and LSTM cells are able to learn longer term dependencies in data. A GRU cell builds on the simple RNN cell by including two additional "gates." These update ($u_t$) and reset ($r_t$) gates in the GRU cell as shown in Figure 4(a) use *sigmoid* activation functions ($\sigma$) and learn which data in a sequence is important to learn and preserve. This ensures that only relevant information is passed along and used to update learnable parameters during the processing of long sequences. LSTMs have more complex cells with multiple gates for learning long-term dependencies as shown in Figure 4(a). An LSTM has three prominent gates, the input gate ($i_t$), the forget gate ($f_t$), and the output gate ($o_t$). These gates utilize the *sigmoid* activation function ($\sigma$). The forget gate acts similarly to the reset gate in a GRU, enabling the model to forget unimportant dependencies in data sequences. The outputs from the forget and input gates are used to generate the LSTM cell state. The cell state ($s_t$), carries relevant long-term dependencies, and uses the *tanh* activation function. Accelerating neural networks that may utilize any of the three types of RNN cells (simple RNN, GRU, LSTM) is essential for devising an efficient sequence learning accelerator. However, it is also challenging due to the diversity across the three RNN types and various connectivity alternatives available in sequence learning models.

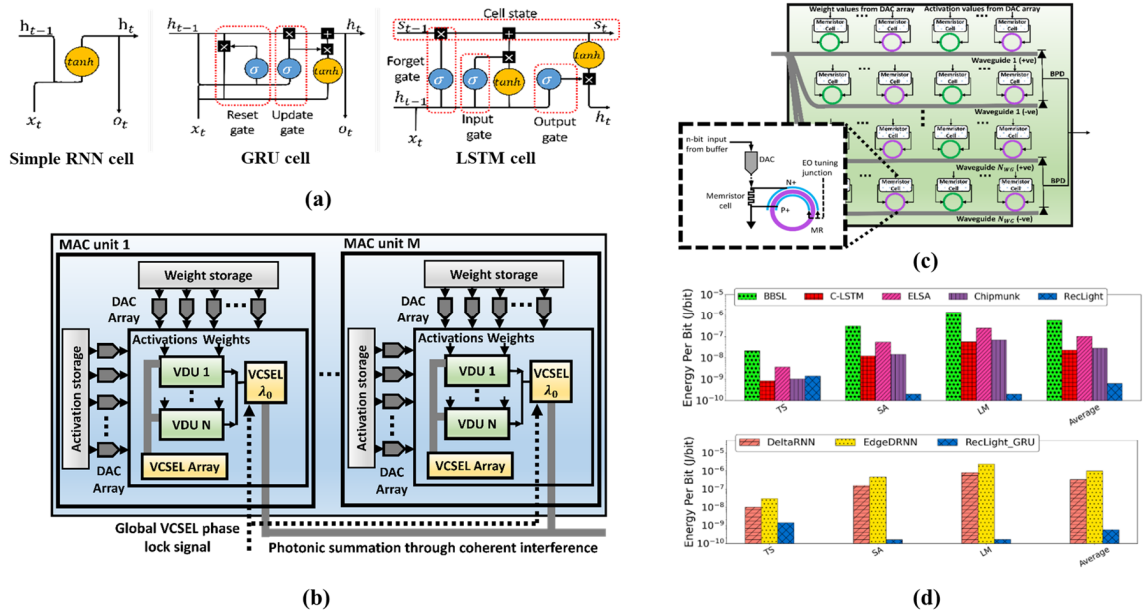

Figure 4. (a) Differences among the fundamental cells in RNN variants: Simple RNN cell, GRU cell; and LSTM cell. (b) A MAC array comprised of MAC units, each with multiple VDUs. Each MAC unit has a vertical cavity surface-emission laser (VCSEL) array driven using the output from the VDU array. (c) VDU showing an MR bank with memristor cells for local parameter storage (BPD: Balanced photodetector). Inset: EO tuning control for memristor cell in VDU. (d) EPB comparison between RecLight and state-of-the-art LSTM accelerators (top) and GRU accelerators (bottom). TS = time series, SA = Sentiment analysis, and LM = language modeling. From [105].

In [105], the first noncoherent optical accelerator for RNNs was presented. The proposed RecLight accelerator leveraged the device, circuit, and architecture level enhancements proposed in CrossLight and made several new innovations to target the acceleration of all three types of RNN models. The RecLight accelerator consisted of multiple MAC units (Figure 4(b)) connected together, where each MAC unit consisted of multiple Vector-Dot-product Units (VDUs; Figure 4(c)). To reduce the power consumption in the DACs, a local parameter storage mechanism was integrated within the VDU that relies on memristors. A memristor cell was integrated into the EO tuning mechanism of an MRR. The conductance of the memristor altered the biasing voltage applied across the EO tuning junction in the MRR. This conductance can be tuned with an appropriate signal from the DAC, and as the memristor can hold this conductance value once the voltage across it is removed, it becomes possible to reuse the same DAC array to tune multiple MRR banks. The banks of MRRs within the VDUs perform the dot-product operations within the RNN cells mapped to them. These banks can also be tasked with accelerating fully connected (FC) layers which usually come after the RNN layers in deep RNN models used in many sequence-learning applications. To support both positive and negative values of parameters involved, separate positive and negative parameter arms are implemented in a VDU, as can be seen in Figure 4(c), for the same waveguide. The sum obtained from the negative arm is subtracted from the sum from the positive arm using a balanced photodetector (BPD) arrangement. The VDUs in a MAC unit share the laser source (vertical cavity surface-emission lasers (VCSELs)) and the DAC array between them (Figure 4(b)), allowing for reuse of the same wavelengths across multiple VDUs, which minimizes the number of VCSELs required and also minimizes laser power consumption. Coherent photonic summation was used to combine the partial sums generated by the MAC units. The non-linear activation functions (sigmoid and tanh) required in RNN cells were implemented using an all-optical approach, with a reconfigurable MRR combined with two Semiconductor-Optical Amplifiers (SOAs), with the ability to be reconfigured to implement both sigmoid and tanh. Figure 4(d) shows the energy per bit comparisons between RecLight and state-of-the-art electronic LSTM accelerators (BBSL, C-LSTM, ELSA, and Chipmunk), and electronic GRU accelerators (DeltaRNN and EdgeDRNN). On average, RecLight obtains at least 37× lower EPB compared to the LSTM accelerators and at least 570× lower EPB compared to the GRU accelerators. The savings are a direct consequence of the device, circuit, and architecture level optimizations integrated into RecLight, the inherent low latency operation of the optical substrate, power optimization to minimize VCSELs and DACs, and reliance on an all-optical non-linearity implementation to minimize E/O/E conversions and their associated latency.

**4.4 Accelerating Transformers**

Transformer neural networks are increasingly being used as part of large language models (LLMs) as well as various other natural language processing (NLP) and computer vision applications. Transformer models enable much higher parallelization than RNNs for sequence modeling and transduction problems, due to their reliance on attention mechanisms and positional encodings instead of recurrence. Since the introduction of the first transformer in 2017 [106], many powerful transformer-based pre-trained NLP models have emerged such as BERT and Albert, along with transformer-based computer vision models, such as the Vision Transformer. Figure 5(a) shows the original transformer model [106] designed for sequence learning. It consists of an encoder that maps a given input sequence into an abstract continuous representation, and a decoder which then processes that representation and gradually produces a single output while also being fed the previous outputs. Input sequences are mapped to a vector, and positional encoding is used to embed the position information of each vector, before being sent to the encoder. The encoder and decoder blocks consist of N stacked layers (Figure 5(a)), with each encoder/decoder consisting of multi-head attention (MHA) and feed forward (FF) layers, along with residual connections for each, followed by layer normalization. The self-attention blocks in the MHA link each element to other elements in a sequence. Each MHA has H self-attention heads, and each attention head generates the query (Q), key (K), and value (V) vectors to compute the scaled dot-product attention. The massive complexity of this transformer model creates many challenges for its acceleration.

In [107], the first optical computing accelerator for transformer neural networks was presented. The proposed TRON architecture used noncoherent optical computing, together with the cross-layer design principles first proposed in Crosslight, power optimizations proposed in RecLight, and new innovations to target the acceleration of a broad family of transformer models. Figure 5(b) shows an overview of the TRON architecture, which is uniquely designed to target the computational needs of Transformer building blocks. The architecture consists of multiple MHA and FF units that can be reused for encoder and decoder blocks. Interfacing with the main memory, buffering of the intermediate results, and mapping the matrices to the optical computing architecture, are handled by an integrated electronic-control unit (ECU).

Unique to the TRON architecture are several optimizations across the design layers. At the device layer, a design space exploration was performed to reduce thermal and crosstalk noise in MRR banks, by exploring combinations of channel spacing (CS) and Q-factor of MRR devices to maximize the tunable range ($R_{tune}$) of these devices while meeting signal-to-noise ratio (SNR) constraints. Figure 5(c) (right) shows the result of this exploration, with the pink star highlighting the best configuration. At the architecture level, several decomposition and co-design enhancements were proposed to efficiently generate and perform operations with the query (Q), key (K), and value (V) vectors. A new optoelectronic circuit was proposed to perform softmax operations and the all-optical non-linear activation implementation from RecLight was extended to support the ReLU and GeLU activation functions. Figure 5(c) (left) shows the results of architectural-level exploration to determine the H (the number of heads in the MHA unit), L (the number of layers), K (the number of rows), and N (the number of columns in each MR bank array) in the optical accelerator that would lead to the lowest EPB/GOPS, where EPB is energy-per-bit and GOPS is giga-operations-per-second. The best configuration with the lowest EPB/GOPS is shown in Figure 5(c) (left) with the pink star. Figure 5(d) shows a comparison of throughput (in GOPS) and EPB for the TRON accelerator, compared to various CPU, GPU, TPU, ASIC, FPGA-based, and resistive in-memory based accelerators for transformer neural networks. The results are shown for four different transformer model variants (x-axis). TRON is able to achieve at least 14× higher GOPS and at least 8× lower EPB than the state-of-the-art, by virtue of its various cross-layer optimizations.

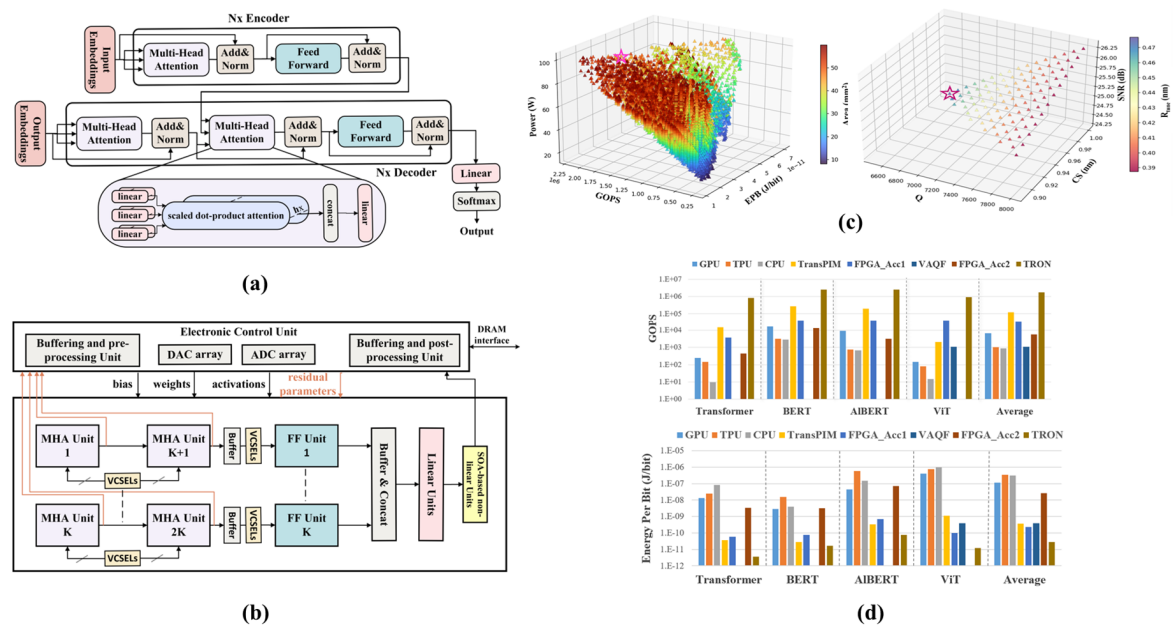

Figure 5. (a) An overview of the transformer neural network architecture. (b) Proposed TRON optical accelerator for transformer neural networks. (c) Architectural optimization for TRON, to find the best [H, L, K, N] configuration with lowest energy and highest throughput (left), where the best configuration is shown with the pink star; MR bank optimization for TRON (right), to identify the best [Rtune,Q,SNR,CS], where the best design point with the highest Rtune is shown with the pink star. (d) Comparison of throughput (top) and EPB (bottom) for TRON with state-of-the-art transformer accelerators. From [107].

## 4.5 Accelerating GNNs

Graph neural networks (GNNs) represent a powerful approach for modeling and learning from graph-structured data. GNNs have been successfully employed across various problem areas, such as recommendation systems, social network analysis, drug discovery, and robotics. GNNs take edge, vertex, and graph feature vectors from input graph structured data and use an iterative approach to process them. Figure 6(a) shows an example of processing with GNNs. The first (aggregation) phase iteratively gathers the neighbors of each vertex and then reduces all data into a single vector. The next phase (reduce operation) can involve a variety of arithmetic functions, e.g., summation, mean, or maximum. This vector is then passed to the final (combination) phase, which usually involves a neural network. Unlike conventional DNNs, where each layer has a different set of weights, vertices in a GNN all share the same weights. GNNs can be composed of several layers, with each layer composed of the phases discussed previously. An iterative process in each layer updates every edge and vertex with information received from immediate neighbor vertices. Thus, relationships with nodes and edges that are progressively farther away can be gradually considered as more layers are processed. Since GNNs were first proposed [108], several variants have emerged, including Graph Convolution Networks (GCNs) that expand the idea of convolution to the graph space and Graph Attention Networks (GATs) which integrate the attention mechanism to determine the usefulness of graph edges more efficiently. Accelerating and efficiently processing GNNs is a significant challenge due to the substantial computational and memory requirements, as well as diversity of GNNs.

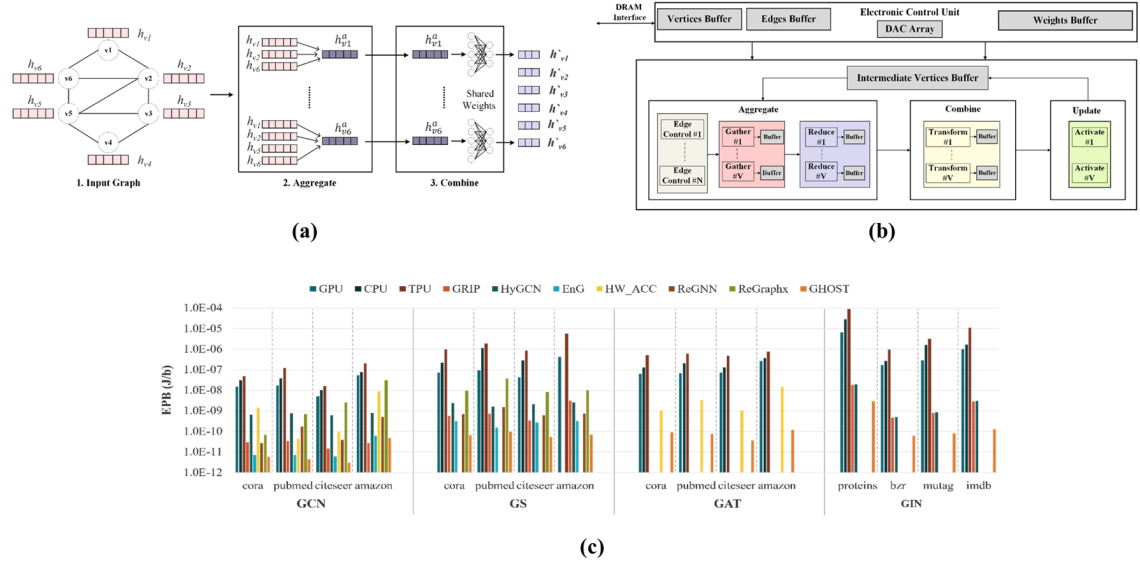

Figure 6. (a) An example of GNN inference showing the input graph to be processed, the Aggregation phase where each vertex's neighbors are reduced to one feature vector, and the Combine and Update phases, where each vertex is linearly transformed and updated using a non-linear activation function. (b) Overview of GHOST accelerator showing the Aggregate, Combine, and Update blocks. (c) EPB comparison between GPU, CPU, TPU, GNN hardware accelerators and GHOST. From [109].

In [109], the first optical accelerator for GNNs was presented. The proposed noncoherent GHOST accelerator targets diverse GNN models, and an overview of the architecture is shown in Figure 6(b). The accelerator is composed of aggregate, combine, and update blocks, enabling the execution of a wide range of GNN models and real-world graph datasets. The accelerator integrated many of the cross-layer reliability, power, and performance innovations from the CrossLight, RecLight, and TRON accelerators. Additionally, GHOST includes several new improvements. At the device level, a more comprehensive approach to overcome heterodyne and homodyne crosstalk was proposed. At the architecture level, unique designs for the aggregate, combine, and update phases were proposed to meet the complex computational needs of these phases. For instance, to support reconfigurable reduction operations in the aggregate phase, such as mean or max, optical comparator devices were integrated with conventional MRR devices. Similarly, in the combine block, Batch Normalization (BN) operations were implemented optically using broadband MRR devices. In the update phase, various activation function types were supported using all-optical MRR and SOA devices, as well as look up tables and simple digital circuits for more complex non-linearity functions. At the software level, four optimizations for efficient orchestration and scheduling of GNNs were proposed, which include techniques for graph buffering and partitioning, execution pipelining and scheduling, weight DACs sharing, and workload balancing. Figure 6(c) shows the EPB results when GHOST was compared with various state-of-the-art GNN hardware acceleration platforms, including CPUs, GPUs, TPUs, and various electronic and memristive GNN platforms. Four different GNN models were considered (GCN,

GraphSAGE, GIN, and GAT) and each of the models was executed with four different node-classification graph datasets (Cora, PubMed, Citeseer, Amazon). On average, GHOST is able to achieve at least 11.1× reduction in EPB compared to the best performing state-of-the-art platform for GNN acceleration.

## 5 HW/SW CODESIGN WITH OPTICAL COMPUTING FOR DNNS IN EDGE/IOT ENVIRONMENTS

Edge and IoT platforms are increasingly being used to host AI applications, e.g., for object detection in automotive embedded perception platforms and IP surveillance cameras. Such platforms typically have stringent constraints on size (form factor), power dissipation, processing capabilities, and memory. As a result, executing DNNs on these platforms while meeting performance goals remains a significant challenge. Optical computing typically has high power overheads which makes it challenging to accelerate DNN execution with an optical substrate. In this section we present some promising solutions to adapting optical computing for edge and IoT platforms.

### 5.1 Power-Aware Design Space Synthesis

To meet the stringent power constraints in edge and IoT platforms, tools are needed for effective design space exploration that can modify and adapt optical accelerators to such constrained environments. In [107], a framework was proposed to enable the deployment of an optical accelerator for transformer neural networks to edge environments. The baseline TRON optical accelerator proposed in [107] had a power budget of 100W which is far too excessive for power budgets in edge environments. The proposed framework constrained the power dissipation to 10W for edge environments and explored the optimal edge configuration values for [H,L,K,N] for implementation on the optical computing platform, where H is the number of implemented heads in the MHA units, L is the number of implemented layers, K is the number of implemented rows, and N is the number of implemented columns in each MR bank array. For the baseline optical accelerator with a 100W power budget, the optimal values that had been derived were [4,2,51,17]. For the edge-friendly optical accelerator with a 10W power budget, the optimal values obtained were [4,1,12,12], with the lower values highlighting fewer implemented components, to meet the more stringent edge power constraints.

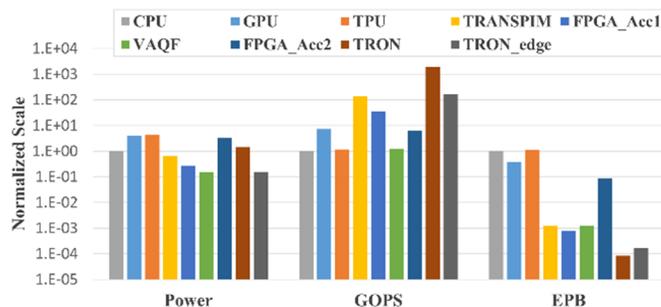

Figure 7. Comparison of power consumption, GOPS, and EPB for the baseline TRON optical accelerator, the edge friendly TRON_edge optical accelerator, and other electronic accelerators for transformer neural networks.

Figure 7 shows the average power dissipation, performance (in terms of GOPS), and energy efficiency (in terms of EPB) across the baseline TRON optical accelerator, the edge friendly TRON_edge optical accelerator, and electronic accelerator for transformer neural networks from prior work. The values shown are normalized to those obtained for the CPU. The synthesized TRON_edge optical accelerator consumes on average much lower power (~10W) than TRON (~100W). While the GOPS values for the edge configuration slightly decrease, its throughput still outperforms all electronic platforms by at least 16%. The EPB value for TRON_edge is higher than for TRON but it still is on average 4× to 6292× lower than the other platforms. The framework from [107] is able to customize the TRON optical accelerator to provide the best performance and energy-efficiency for any given target power consumption constraint.

### 5.2 Pruning Approaches

An interesting approach to reduce the performance and power overheads of CNN model execution is to sparsify the DNN models, i.e., reduce carefully selected model parameters to zero during or after training while maintaining accuracy goals, to reduce the computations required during execution. But simply deploying such sparse CNNs on an accelerator does not necessarily ensure model performance and energy-efficiency improvements. This is because most accelerators today are optimized for executing dense neural networks and are not able to take advantage of the sparsity available in neural networks. Dense neural network accelerators inefficiently end up orchestrating dataflow and operations for parameters that have been sparsified (i.e., zeroed out). By having to unnecessarily process sparse parameters, these conventional accelerators result in high latency and high energy consumption when sparse neural networks are executed on them. Therefore, strategies for taking advantage of sparsity and reducing the number of operations are essential in hardware designed to execute sparse neural networks.

In [110], a novel sparse neural network optical accelerator was proposed. The proposed SONIC accelerator employed sparsity-aware data compression and dataflow techniques for fully connected and convolution layers, which were tuned for high throughput operation on the optical acceleration substrate. The hardware designs were also modified to efficiently operate with sparse parameters, e.g., vector dot product units were updated to efficiently work with sparse vectors and power gating mechanisms were employed to reduce power dissipation across optical devices involved in the computations. Figure 8 compares the EPB for SONIC against various optical and electronic CNN accelerators. SONIC exhibits at least 5.78× lower EPB than the most efficient platform from prior work, showcasing the impact of optimizing the optical acceleration platform for model sparsity.

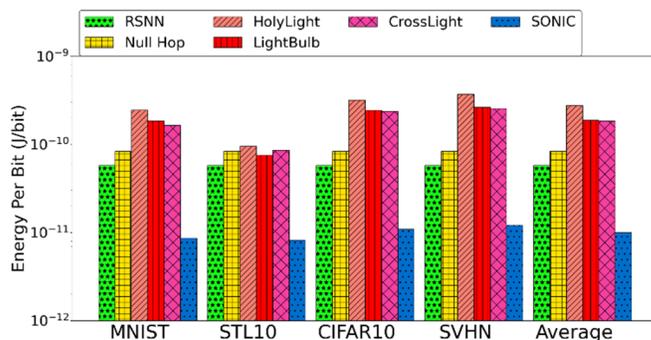

Figure 8. The EPB of the proposed SONIC optical accelerator compared to various optical accelerators (CrossLight, HolyLight, LightBulb) and electronic accelerators (RSNN, Null Hop) for CNNs. From [110].

### 5.3 Quantization Approaches

Another approach to realize optical accelerators for power and performance constrained edge and IoT platforms is to utilize quantization in DNNs and perform hardware/software co-design to devise an optical accelerator tuned for the chosen quantization level. In quantization, the bitwidths of DNN parameters (weights) and activations are reduced from the default (typically 32-bit floating point) to a lower value (e.g., 8 bit fixed-point). In [111], an optical accelerator for binary quantized CNNs was proposed. Here binary refers to utilizing a single bit to represent parameters and/or weights. Unfortunately, using such a small bitwidth significantly reduces CNN inference accuracy. Studies have shown that activations are more susceptible to bitwidth reduction than weight parameters. Therefore, an alternative for increasing inference accuracy is to not binarize activations, while keeping the weights binarized. To determine the appropriate activation bitwidth (precision), an analysis was performed in [111], where weights were restricted to binary (1-bit) values, but the bit precision level of the activations was altered from 1-bit to 16-bits, across various CNN models and learning tasks. During CNN model training, weights were binarized during the forward and backward propagations but not during the parameter update step, because keeping high precision weights during the updates is necessary for stochastic gradient descent (SGD) to work efficiently. After training, all weights were in binary format, while the precision of input activations was varied. Figure 9(a) shows the results of varying activation bitwidth across four different CNN models and their datasets. It was observed that the accuracy had notable change initially as activations bits were increased, but this gain in accuracy soon saturated. Based on the results, 1-bit weights were considered along with 4-bit activations, which allowed reducing bitwidths without significantly reducing model accuracy.

The reduction in bit-width for weights and activations, from a baseline of 32-bit to 1 and 4 bits, respectively, allowed significantly reducing the number of ADCs and DACs in the proposed binary optical accelerator called ROBIN [111]. The use of such low bitwidth weights and activations also allowed simplifying the optical computing architecture design by reducing the number of photonic devices required. These low bitwidth driven modifications allowed for a significant reduction in power dissipation. Two variants of the proposed ROBIN optical accelerator for binarized CNNs were proposed. The first was optimized for maximizing frames per second per watt (FPS/Watt), with lowest area and power consumption (called the energy optimized ROBIN or ROBIN-EO variant), and another with the best FPS but with higher area and power consumption (called the performance optimized ROBIN or ROBIN-PO variant).

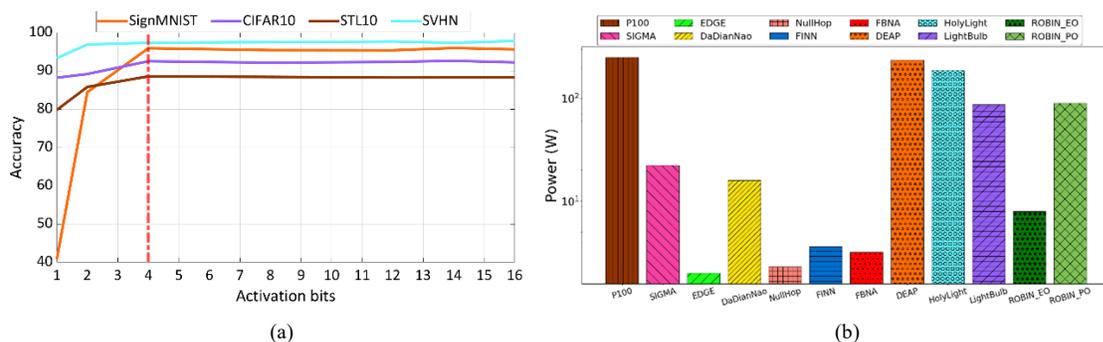

Figure 9. (a) The accuracy sensitivity study conducted by varying activation parameter precision (number of bits). Weights are kept as binary values in all cases. The study was performed across four different models and their datasets. (b) Power dissipation among two variants of the proposed ROBIN accelerator and different optical CNN accelerators (DEAP-CNN, Holylight, LightBulb), and electronic CNN accelerator platforms (P100, SIGMA, EdgeTPU, DaDianNao, Null Hop, FINN, and FBNA). ROBIN_EO is an energy optimal configuration of the proposed binary optical neural network accelerator, while ROBIN_PO is a performance (throughput) optimized version of the proposed binary optical neural network accelerator. From [111].

Figure 9(b) compares the power dissipation of the two variants of the proposed ROBIN architecture with various optical and electronic CNN accelerators. It can be observed that ROBIN-PO has substantially higher power consumption than ROBIN-EO, as ROBIN-PO is focused on FPS performance rather than energy conservation. ROBIN-PO has a much larger vector granularity per VDP unit along with substantially higher VDP unit count to maximize parallelism, when compared to ROBIN-EO. The larger unit count and the waveguide count in ROBIN-PO increases its power overhead. In contrast, it can be observed that the energy and area efficient ROBIN-EO has comparable (less than 10 W) power consumption to that of edge and IoT electronic neural network accelerators such as EDGE, FINN, FBNA, and NullHop. ROBIN_EO was also shown to have the lowest EPB compared to all optical and electronic accelerators that were compared against it.

In [112] a heterogeneous quantization based optical accelerator for CNNs was proposed. Conventional quantization approaches use the same bitwidth for all the weight and activation parameters across layers, which is referred to as homogeneous quantization. In the heterogeneous or mixed precision quantization approach used in [112], different layers in a DNN were allowed to have different levels of quantization to achieve lower memory and computational complexity, while maintaining model accuracy. The proposed approach leveraged the key insight that certain CNN layers are more critical for maintaining high inference accuracy than others. Therefore, if the bitwidth of the critical layers is kept higher than that of other layers, during quantization, it can help improve model accuracy.

The proposed HQNNA optical accelerator in [112] was designed to support the execution of heterogeneously quantized CNNs. It made use of innovative time division multiplexing (TDM) and bit-slicing approaches to aggressively reduce energy consumption while supporting optical operations with heterogeneous bitwidths. Figure 10 shows the EPB for HQNNA compared to various other optical accelerators, including CrossLight and ROBIN, that were discussed earlier. The results are shown across multiple CNN models with similar accuracy, where HQNNA was used to efficiently execute heterogeneously quantized CNNs, while the other accelerators executed homogeneously quantized versions of the same CNN models. It can be observed that heterogeneous quantization, together with the hardware adaptations to support it in HQNNA allows the optical accelerator to achieve much lower EPB than all state-of-the-art optical CNN accelerators.

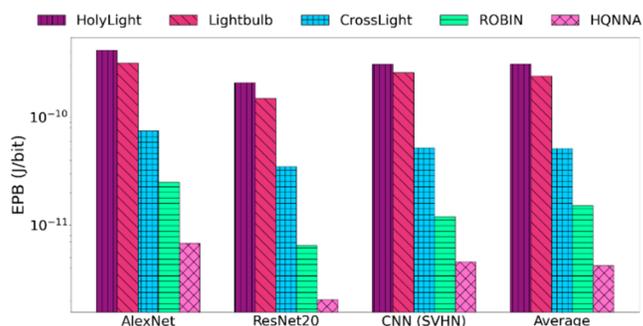

Figure 10. EPB comparison across three different CNNs for HQNNA (which supports heterogeneous quantization) and other optical CNN accelerators that support homogeneous quantization. From [112].

## 6 OPEN PROBLEMS AND FUTURE DIRECTIONS WITH OPTICAL DNN ACCELERATION

Clearly there is a tremendous momentum with exploring and adapting optical computing for accelerating DNNs, which is arguably the most pressing computational challenge facing modern computer hardware design. Great strides have been made to overcome integration, interfacing, and software infrastructure challenges with optical computing in recent years. Several startup companies have emerged that have begun to adopt optical computing for DNN acceleration in commercial products [113]-[116]. To continue to adapt and meet the needs of emerging DNNs, several open problems exist that can allow us to chart a roadmap for future directions in this vibrant field. Some of the key directions are discussed below.

- **Tolerating Variations:** Integrated optical devices such as MRRs, MZIs, and waveguides used in optical DNN acceleration platforms typically require post-manufacture correction for semiconductor fabrication variations and environmental sensitivity via passive trimming mechanisms, such as electron beam induced compaction and strain of oxide cladding [117] and the CMOS foundry compatible approach of germanium ion implantation and annealing [118]. Beyond these design time variations, thermal variations due to temperature changes induced at runtime can also cause resonance drifts in resonant device such as MRRs, as discussed in Section 3.2, which is problematic as it decreases SNR and increases errors during computations and communication on the optical substrate. Active trimming can counteract such runtime thermal variations, for instance by changing the refractive index of a waveguide via integrated heaters. But this approach requires constant and high input power for the heater, and also slows down the effective throughput achievable with the devices due to associated microsecond response speeds. Electro-optic mechanisms are faster, but their correction range is limited. While these electro-optic mechanisms can be deployed in combination with heater-based thermo-optic mechanisms (as discussed in Section 4), the approach can still lead to slowdowns when large resonance drifts occur at runtime. Thus, new approaches are needed with fast response and large correction range to cope with runtime variations. One promising direction is to utilize cross-layer optimizations to manage temperature on optical DNN accelerator platforms via intelligent mapping and load balancing of computations, as discussed for photonic chip-scale networks in [119], [120]. New mechanisms are also required for efficient management of homodyne and heterodyne crosstalk variations in optical DNN platforms. Approaches proposed for minimizing runtime crosstalk in photonic chip-scale network platforms, e.g., [121]-[125], could be adapted for optical DNN acceleration platforms.

- **Efficient Optoelectronic Conversions:** Optical computing platforms need to interface with digital electronic components, which requires using costly conversion steps with DAC and ADC circuits operating at full data rate (tens of gigahertz). New approaches are needed to reduce the complexity and energy overheads of these conversion steps. A promising solution involves photonic DACs that can combine digital to analog conversion and electro-optic modulation. Implementations of this approach using optical intensity weighting of multiwavelength signal, modulated with silicon MRRs to realize 2-bit photonic DACs were presented in [126], [127]. Using a coherent parallel photonic DAC can allow for scaling to higher numbers of bits [128]. Such photonic DACs can reduce area and power consumption in optical DNN accelerators while supporting high sampling rates, high precision, and low distortion.

- **Optical Storage:** Optical computing platforms are tightly coupled with electronic platforms which provide general purpose computing capabilities with logic gates and memory. The reliance on electronic memory (and associated conversion overheads) can be minimized by realizing data storage in the optical domain. This idea has been recently explored by storing neuron weights in DNNs within phase change materials (PCMs) integrated into both noncoherent optical DNN accelerators [105] and coherent optical DNN accelerators [129]. In these implementations, precomputed weights for a given DNN inference task can be stored in PCMs integrated within the optical computing components (such as MRRs, waveguides) to enable efficient real-time inference for any given input (activations) to the DNN. Commonly used PCMs such as $Ge_2Sb_2Te_5$ and low-loss GeSbSeTe possess multiple stable phases of matter that have distinct optical properties. It is possible to vary the refractive index of photonic waveguides, MRRs, and MZIs with integrated PCM, by changing the state of the PCM from amorphous to crystallize either via electronic heating or optically with lasers [130]. A recent work also showed how optical main memory based on PCMs can be realized [131]. By integrating such optically controlled memories with optical DNN accelerators, it can be possible to improve the throughput and energy efficiency of future optical DNN accelerator platforms.

- **Scalability:** Optical losses limit the scalability and achievable accuracy in optical DNN accelerator platforms. Losses during coupling of optical signals from external sources, as well as within devices used in optical computations (such as directional couplers and phase shifters) cause attenuations in optical signal intensities reaching the outputs [132], which creates severe challenges when scaling the designs of optical DNN accelerators. Specifically, the optical intensities required at the photodetectors to detect an optical signal with a desired resolution (in terms of bits) places a limit on the size of the matrix that can be computed optically [132]. A recent study also showed how optical losses contribute to a significant 84% drop in inference accuracy in coherent (MZI mesh based) optical DNN accelerators [133]. New mechanisms are thus needed to reduce losses, to improve accuracy and scalability in optical computing. Recent promising developments with photonic wire bonds (PWBs), which involves writing three-dimensional waveguides in a photosensitive polymer, have demonstrated efficient interfacing between the external sources to the silicon waveguide with very low coupling losses [134]. Optical signal attenuation due to losses can also be compensated by using SOAs that can amplify their input signals. However, improvements in the design of SOAs are needed, as current SOAs have high energy overheads and the non-linear gain-current curve in these SOAs also necessitates careful calibration. Increasing laser power can also be used to compensate for losses, but this approach negatively impacts energy efficiency. Approaches to adaptively reconfigure laser power [135] to opportunistically increase it when needed for loss compensation and lowering it when compensation is unnecessary can be useful in this context. Improving the design and responsivity of photodetectors is another way to tolerate losses without increasing laser power, e.g., by utilizing avalanche photodetectors [136] and carefully designing their geometry to improve dark current and quantum efficiency, and using equalization techniques [137] can improve the sensitivity of these devices to losses. The use of low-loss silicon nitride waveguides can also help reduce signal propagation losses [138]. Lastly, employing 2.5D integration to design an optical DNN accelerator in a modular manner, as discussed in [139], [140] can also improve scalability and allow the mapping of larger DNNs onto the optical computing substrate.
- **DNN Inference vs. Training:** The DNN accelerators proposed to date across industry and academia focus primarily on accelerating DNN inference. Training DNNs to update their weight parameters for a given learning task is primarily based on backpropagation of error gradients. This requires calculating the error gradients for each network connection in the DNN according to the chain rule of differentiation. In the optical domain, this requires implementing an additional backward path where each weight parameter needs to be probed and updated, which has prohibitively high complexity. Gradient calculation also relies on electronic circuits, and the required electro-optic conversions for these would further slow down each iteration of the backpropagation phase. Further, to optically implement weight updates according to the error backpropagation approach, a neuron's nonlinearity in the backward direction needs to be the gradient of its nonlinearity in the forward direction. Such asymmetric neurons are a challenge to realize optically. Thus, implementing training efficiently in the optical domain remains a significant open challenge. Several recent efforts are attempting to overcome these drawbacks by avoiding error backpropagation altogether, including using equilibrium propagation [141] which adjusts weights using local contrastive Hebbian learning, employing statistical optimization tools such as genetic algorithms or Bayesian optimization [142], and utilizing coordinate descent approaches where sets of weights (i.e., coordinates) are selected at random, modified to probe the error landscape's local gradient, and then updating the weights in a direction that is opposite to the gradient's direction [143]. Another approach involves the use of an adjoint variable method to implement a photonic analogue of the backpropagation algorithm, as outlined in [144] for coherent optical DNN implementations. In this approach, the gradient of the loss function with respect to each phase shifter in a coherent MZI array can be obtained simultaneously by physically propagating the original field, the adjoint field, and the interference field. While such an approach for gradient measurements was shown to be exact in a lossless system, its effectiveness in practical (lossy) platforms and applicability to more complex DNN training problems remains an open problem.

- **Security:** Optical DNN accelerators are susceptible to various attacks, much like their digital electronic counterparts. Examples of these attacks include data confidentiality attacks that steal information, data integrity attacks that corrupt information, and denial of service attacks that render a platform unusable by the intended users. The modality by which practical variants of these attacks can be launched in optical DNN accelerator platforms remains an open problem, as do techniques that can mitigate their impact. As these platforms proliferate, it is crucial to characterize and be able to overcome such attacks, to build trust among the user community and prevent catastrophic failures. In [145]-[147] some of the first approaches to realize data confidentiality attacks in chip-scale photonic platforms were proposed, along with mechanisms to detect and counter such attacks. Similar approaches must be devised for emerging optical DNN accelerator platforms.

# 7 CONCLUSION

This article surveyed the historical trends, recent developments, and open challenges with optical computing for DNN acceleration. Optical DNN acceleration has many promising advantages over conventional digital electronic DNN acceleration platforms. Optical multiply and accumulate (MAC) operations can be scaled to frequencies up to tens of GHz, unlike MAC operations in digital electronic platforms with a few hundreds of MHz or at most a few GHz operating speeds. Optical domain multiplications with matrix/vector data also have significantly higher throughput and energy efficiency than digital platforms, as these operations can be intrinsically implemented in parallel, where the analog nature of the computation allows all matrix operations to take place concurrently. These fundamental benefits with optical computing have invigorated interest in realizing practical optical DNN accelerator systems, with several startups having already showcased proof of concept prototypes that outperform electronic counterparts. Nonetheless, many open challenges remain, related to practical concerns with scalability, reliability, security, training efficiency, and overheads due to reliance on electronic components, which must all be addressed to allow the field of optical DNN acceleration to evolve to meet the needs of increasingly complex DNNs that are powering emerging applications that impact every sphere of our lives.